\documentclass[twocolumn,floatfix, nofootinbib,prd,preprintnumbers,showpacs,showkeys,superscriptaddress,longbibliography]{revtex4-2}
\usepackage{slashed}
\usepackage{amsmath}
\usepackage{booktabs}
\usepackage{array}

\usepackage{graphicx} 
\usepackage{orcidlink}
\usepackage[top=2cm,bottom=2cm,left=1cm,right=1cm,marginparwidth=1.75cm]{geometry}
\usepackage{mathrsfs}
\usepackage{amssymb}
\usepackage{amsmath}
\date{\today}
\usepackage{mathtools}
\usepackage{comment}

\usepackage[mathscr]{euscript}
\usepackage{amsmath}
\usepackage{graphicx}
\usepackage{dcolumn}
\usepackage{bm}
\usepackage{epsfig}
\usepackage{amssymb,latexsym,mathrsfs}
\usepackage{graphicx}
\usepackage{color}
\usepackage{hyperref}
\usepackage{float}
\usepackage{diagbox}

\usepackage{xcolor}
\definecolor{darkblue}{RGB}{0,0,150}

\usepackage{textgreek}

\usepackage{tikz}

\usepackage{cellspace}
\usepackage{amsmath}
\usepackage{physics}
\usepackage{csquotes} 
\newcommand*\diff{\mathop{}\!\mathrm{d}}

\hypersetup{
    colorlinks=true,
    linkcolor=red,
    citecolor=blue,
} 

\usepackage{tcolorbox}
\usepackage{amsmath}
\usepackage{amssymb}
\usepackage{subfigure}
\usepackage{hyperref}
\usepackage{url}
\usepackage{xcolor}
\usepackage{color}
\usepackage{soul}
\definecolor{amaranth}{rgb}{0.9, 0.17, 0.31}
\definecolor{purple(munsell)}{rgb}{0.62, 0.0, 0.77}
\definecolor{americanrose}{rgb}{1.0, 0.01, 0.24}
\definecolor{palatinateblue}{rgb}{0.15, 0.23, 0.89}
\definecolor{royalblue(web)}{rgb}{0.25, 0.41, 0.88}
\definecolor{hanpurple}{rgb}{0.32, 0.09, 0.98}
\definecolor{beaublue}{rgb}{0.74, 0.83, 0.9}
\definecolor{carminered}{rgb}{1.0, 0.0, 0.22}
\definecolor{brightpink}{rgb}{1.0, 0.0, 0.5}
\definecolor{vividviolet}{rgb}{0.62, 0.0, 1.0}
\hypersetup{ linktoc=all,
    colorlinks, linkcolor={palatinateblue},
    citecolor={brightpink}, urlcolor={amaranth}}

\usepackage{comment}

\newcommand{\be}{\begin{equation}}
\newcommand{\ee}{\end{equation}}
\newcommand{\bs}{\begin{split}} 
\newcommand{\bea}{\begin{eqnarray}}
\newcommand{\eea}{\end{eqnarray}}

\newcommand{\bes}{\begin{subequations}}
\newcommand{\ees}{\end{subequations}}

\newcommand{\bo}{\raise-1mm\hbox{\Large$\Box$}}

\begin{document}

\preprint{FTPI-MINN-26-12}
\preprint{UMN-TH-4532/26}

\title{Apparent Fermionic Spectra for Bosonic Radiation: Accelerated Charge Kinematics}

\author{Michael R.R. Good\, \orcidlink{0000-0002-0460-1941}}
\email{muon@asu.edu}
\affiliation{Physics Department \& Energetic Cosmos Laboratory, Nazarbayev University,\\
Astana 010000, Qazaqstan.}
\affiliation{Leung Center for Cosmology \& Particle Astrophysics,
National Taiwan University,\\ Taipei 10617, Taiwan.}
\affiliation{Beyond Center for Fundamental Concepts in Science, Arizona State University,\\
Tempe AZ 85287, USA.}

\author{Evgenii Ievlev\,\orcidlink{0000-0002-5935-4706}}
\email{ievle001@umn.edu}
\affiliation{William I. Fine Theoretical Physics Institute, School of Physics \& Astronomy,
University of Minnesota,\\ Minneapolis MN 55455, USA.}

\author{Arsen Almaskhan\, \orcidlink{0009-0008-1875-8449}}
\email{a.almaskhan@spectrum.edu.kz}
\affiliation{Physics Department \& Energetic Cosmos Laboratory, Nazarbayev University,\\
Astana 010000, Qazaqstan.}
\affiliation{Science Department, Spectrum International School,\\ Astana 010000, Qazaqstan.}


\begin{abstract} 
An accelerated point charge can emit photons with an apparent Fermi–Dirac spectrum, even though the radiation is bosonic and its occupation numbers are not constrained to 0 or 1. The effect arises from a special class of acceleration kinematics and does not rely on thermal equilibrium, horizons, or statistical ensembles.

\end{abstract}
\maketitle

\section{Introduction}

A foundational result in statistical physics and quantum field theory is that bosonic excitations follow Bose–Einstein (BE) statistics, whereas fermionic excitations obey the Fermi–Dirac (FD) statistics. Nevertheless, the literature on the moving-mirror \cite{DeWitt:1975ys,Davies:1976hi,Davies:1977yv}, also known as the dynamical Casimir effect (DCE) \cite{moore1970quantum,physics7020010}, contains early hints of a fermionic spectral structure in bosonic radiation. In particular, Haro and Elizalde \cite{Haro:2008zza} identified a FD–type form emerging in the large $\omega'$ frequency behavior of the Bogoliubov coefficient for a semitransparent mirror. This feature was subsequently confirmed by Nicolaevici \cite{Nicolaevici:2009zz} for the energy variable $\omega$, who emphasized that the result described only an asymptotic regime and therefore did not establish a genuine particle-number distribution. Motivated by this observation, Elizalde and Haro \cite{Elizalde:2010zza} called for a deeper analysis of the origin of the fermionic form, its connection to the sign structure of the Bogoliubov coefficients, and its implications for mode occupation.

In a recent work, two of the present authors studied photons exhibiting a Fermi-Dirac angular distribution in the far-field regime and for a particular angular configuration \cite{Ievlev:2023akh}.  The angular-distribution context clarified the subtleties using classical physics, without appealing to the DCE. The interpretation was further strengthened by the exact electron-mirror correspondence \cite{Ford:1982ct,Unruh:1982ic,Ritus:2003wu,Ritus:2002rq,Ritus:1999eu,Nikishov:1995qs,Zhakenuly:2021pfm,Ritus:2022bph,Ievlev:2023inj,ptep,Ievlev:2023xzv}, which relates radiation from accelerated charges in 3+1 dimensions to moving-mirror particle production in 1+1 dimensions.

In contrast to the asymptotic and angular indications \cite{Haro:2008zza,Nicolaevici:2009zz,Elizalde:2010zza,Ievlev:2023akh}, the present work demonstrates the kinematic emergence of a non-relativistic Fermi–Dirac spectrum for bosonic radiation, without the need for a particular angular dependence. 
We investigate the phenomenon using a moving point charge in classical electrodynamics, e.g, treated in texts \cite{Jackson:490457} and \cite{Zangwill:1507229}, showing that no appeal to quantum field theory, semi-transparent boundaries, or horizon thermality is required. 

The paper is organized as follows. Section~\ref{sec:FDspectra} introduces the target Fermi--Dirac photon spectrum and derives the associated number density, total particle count, total emitted energy, and Wien displacement scale. Section~\ref{sec:LLkinematics} constructs the nonrelativistic Larmor--Li\'enard point-charge trajectory that realizes this spectrum, identifies the Lambert-\(W\) Fourier--Mellin structure responsible for the Fermi--Dirac denominator, and shows that the radiation is emitted in a burst-like, out-of-equilibrium time profile. Section~\ref{sec:info} analyzes the spectral information content and the thermodynamic analogy, including Shannon entropy, spectral indistinguishability from an equilibrium Fermi--Dirac distribution, and effective-volume scaling that restores a Stefan--Boltzmann-like \(T^4\) behavior. Finally, Section~\ref{sec:conc} summarizes the central result: bosonic radiation from a classical accelerated charge can exhibit an apparent fermionic spectrum without invoking quantum statistics, thermal equilibrium, or horizons. 

We use natural units, $\hbar = c = k_B = \mu_0 = 1$. The electron charge and fine-structure constant are related by $\alpha = e^2/4\pi$.

\section{Fermi--Dirac Spectra}
\label{sec:FDspectra}
\subsection{Spectrum \& Particles}
Consider the frequency-domain Fermi-Dirac energy spectrum: 
\begin{equation}
    E_{\text{FD}}(\omega) = \hat{E}_0 \frac{4 \alpha }{3} \frac{(\omega/\kappa)^3}{e^{2\pi\omega/\kappa} + 1}.
\label{fd_spectrum_1}
\end{equation}
Here, $\hat{E}_0$ is a free, dimensionless parameter that sets the overall energy scale.
Distribution Eq.~\eqref{fd_spectrum_1} corresponds to a temperature 
\begin{equation}
    T = \frac{\kappa}{2 \pi} \,.
\end{equation}
The number density spectrum $N(\omega)$ is related to the energy spectrum by the semi-classical \cite{Jackson:490457} relation $N(\omega) = E(\omega)/\omega$, yielding:
\begin{equation}
    N(\omega) = \hat{E}_0 \frac{4 \alpha}{3 \kappa}\,\frac{(\omega/\kappa)^2}{e^{2\pi\omega/\kappa} + 1}. \label{eq: number density spectrum}
\end{equation}
Consequently, the total number of particles $N$ is found by integrating over the frequency domain:
\begin{equation}
    N = \int_{0}^{\infty} N(\omega) \, d\omega = \hat{E}_0 \frac{\zeta(3)}{4\pi^{3}}\,\alpha. \label{eq: total number of particles}
\end{equation}

\subsection{Energy \& Wien's Displacement }

The total emitted energy $E$ is obtained by integrating the frequency-weighted number density:
\begin{equation}
    E = \int_{0}^{\infty} \omega N(\omega) \, d\omega 
    = \hat{E}_0 \frac{7\,\zeta(4)}{16\pi^4}\,\alpha \kappa
    = \hat{E}_0 \frac{7 \alpha \kappa}{1440}. \label{eq: total emitted energy}
\end{equation}
Of course, the total energy can be found directly from the Fermi-Dirac spectrum without appeal to the semi-classical count via
\be
E = \int_0^\infty E(\omega) \diff{\omega} = \hat{E}_0 \frac{7\alpha \kappa}{1440}.
\ee 
Consider the value of $\omega$ that maximizes the function $E(\omega)$ 
\begin{equation}
    \frac{dE(\omega)}{d \omega}=0 \rightarrow \frac{\omega_\text{max}}{\kappa}  = \frac{3}{2 \pi}\,(1+e^{-2 \pi \omega_\text{max} / \kappa}) \,.
\end{equation}
Of course, $\omega_{\max}$ scales linearly with $\kappa$, as follows already from dimensional analysis.
Specifically, one finds
\begin{equation}
\frac{\omega_{\max}}{\kappa}
= \frac{3 + W\left(3e^{-3}\right)}{2\pi} \approx 0.498.
\end{equation}
This is just Wien's displacement law, but in our kinematic case, the temperature is related to the acceleration scale $\kappa$, 
\begin{equation} 
\frac{\omega_{\max}}{T} = 3 + W\left(3e^{-3}\right) \approx 3.131.
\end{equation}
This agrees with the fermionic equilibrium result when the acceleration parameter is identified with the temperature via $\kappa = 2\pi T$. For contrast, the bosonic statistical equilibrium yields Wien’s displacement law $\omega_{\max}/T = 3 + W(-3e^{-3}) \approx 2.821$.

\section{Larmor--Liénard Kinematics}
\label{sec:LLkinematics}
In this Section, we provide a trajectory of a classical point charge, such that the resulting spectrum of the radiated electromagnetic waves precisely matches the distribution Eq.~\eqref{fd_spectrum_1}.

\subsection{Kinematics \& Power}

To begin, let us set a few basic equations and fix our conventions.
In our units, the non-relativistic Larmor formula gives \cite{Larmor1897}
\begin{equation}
    P(t) = \frac{2}{3} \alpha \, a^2(t) \,,
\label{powerLL}
\end{equation}
where $a(t) = \ddot{z}(t)$ is the acceleration, and we consider a point charge in a rectilinear motion along the $z$ axis with the time dependence $z(t)$. 

The spectral-angle distribution is found as (see, SI units e.g., Eq. (23.89) p. 911 of Zangwill \cite{Zangwill:1507229}
or Gaussian units Eq. (14.67) p. 701 of Jackson \cite{Jackson:490457})
\begin{equation}
    \frac{\diff I(\omega)}{\diff \Omega} = 
    \frac{\alpha \; \omega^2 \sin^2\theta }{4\pi^2}
    \left|\int\displaylimits_{-\infty}^{\infty} \diff t\, \dot{z}(t) e^{i\phi}\right|^2,
\label{angular_spectrum_basic_formula}
\end{equation}
where $\phi = \omega (t - z(t)\cos\theta)$.
The spectrum $E(\omega)$ is then given by
\begin{equation}
    E(\omega) = \int \diff \Omega  \frac{\diff I(\omega)}{\diff \Omega}.
\end{equation}
Of course, self-consistency guarantees that the total radiated energy, computed via the spectrum or via the power, is the same,
\begin{equation}
    E_{\rm tot}=\int_{0}^{\infty} \diff\omega\;E(\omega)
=\int_{-\infty}^{\infty} \diff t\;P(t),
\label{eq:ParsevalEnergy}
\end{equation}
Note that $E(\omega)$ is not in any way a Fourier transform of $P(t)$.

In this work, we focus on the non-relativistic limit $|\dot{z}| \ll 1$.  In fact, in this limit 
\begin{equation}
    \phi = \omega (t - z(t)\cos\theta) = \omega t + \mathcal{O}( t \, \dot{z}(t) ).
\end{equation}
This allows us to simplify the spectral-angle distribution in Eq.~\eqref{angular_spectrum_basic_formula},
\begin{equation}
    \frac{\diff I(\omega)}{\diff \Omega} \Bigg|_\text{non-rel} 
    \approx \frac{\alpha \; \omega^4 \sin^2\theta }{2\pi} \left| z(\omega) \right|^2.
\label{specdistr_I_non-rel}
\end{equation}
Here, $z(\omega)$ is the Fourier transform (our conventions are summarized in Appendix~\ref{sec:fourier}).
Integrating Eq.~\eqref{specdistr_I_non-rel} over the solid angle, we obtain the spectrum $E(\omega)$,
\begin{equation}
    E(\omega) \Bigg|_\text{non-rel} \approx 
    \frac{4 \alpha }{3} \omega^4
    \left| z(\omega) \right|^2,
\label{E_spectrum_non-rel_leading}
\end{equation}
In this case, the general equality Eq.~\eqref{eq:ParsevalEnergy} reduces to the Parseval-Plancherel theorem, see also \cite{Zangwill:1507229}.

We now seek a trajectory for which the radiated energy spectrum in Eq.~\eqref{E_spectrum_non-rel_leading} matches the Fermi-Dirac distribution in Eq.~\eqref{fd_spectrum_1}.
In principle, there can be many such trajectories.
In this work we report on the following (see Sec.~\ref{subsec:imaginary-time-monodromy} below for the proof):
\begin{equation}
    z(t) = \frac{2 \sqrt{\hat{E}_0} }{ \kappa^2 } \frac{\diff}{\diff t} \sqrt{W(e^{\kappa t})}
    = \frac{ \sqrt{\hat{E}_0} }{ \kappa } \frac{\sqrt{W(e^{\kappa t})}}{ 1 + W(e^{\kappa t}) }.
\label{z_fd_explicit}
\end{equation}
Here, $W$ denotes the Lambert $W$-function, also called the product log.
From the asymptotics 
\begin{equation}
\begin{aligned}
    W\!\left(e^{\kappa t}\right) &\sim e^{\kappa t} \qquad &(t \to -\infty), \\
    W\!\left(e^{\kappa t}\right) &\sim \kappa t - \ln(\kappa t)  \qquad &(t \to +\infty),
\end{aligned}
\end{equation}
we can infer that the trajectory in Eq.~\eqref{z_fd_explicit} asymptotically starts at $z=0$ and asymptotically returns to $z=0$, with vanishing initial and final velocities.
In other words, the trajectory at hand is an asymptotically resting round-trip trajectory, see also a recent study of this family of motions \cite{Mujtaba:2024vmf}.
The maximum distance from the origin in the course of this motion is given by
\begin{equation}
    z_\text{max} = \frac{\sqrt{\hat{E}_0} }{ 2 \kappa }.
\end{equation}
The maximal speed can also be easily found to be
\begin{equation}
    v_\text{max} = \frac{1}{432} \sqrt{ 2 \hat{E}_0 (587+143 \sqrt{13})} \approx 0.109 \sqrt{\hat{E}_0}.
\label{v_max}
\end{equation}
The parameter $ \hat{E}_0 $, therefore, controls the validity of the non-relativistic approximation.

\subsection{The Fermi-Dirac denominator and the Imaginary-time monodromy}
\label{subsec:imaginary-time-monodromy}

In fact, a more general class of trajectories can also yield a spectral distribution with the same denominator, in the Fermi-Dirac form as in Eq.~\eqref{fd_spectrum_1}.
To see how this comes about, consider the
family
\begin{equation}
    z_p(t)
    =
    C_p\frac{d}{dt}
    \left[W(e^{\kappa t})\right]^p,
    \qquad p>0 .
    \label{eq:zp-family}
\end{equation}
The trajectory in Eq.~\eqref{z_fd_explicit} corresponds to
\(p=1/2\), with
\begin{equation}
    C_{1/2}=\frac{2\sqrt{\hat E_0}}{\kappa^2}.
\end{equation}

Now, let us compute the Fourier transform of the trajectory.
To this end, consider a natural Lambert-\(W\) variable
\begin{equation}
    u \equiv W(e^{\kappa t}), \qquad
    u e^u = e^{\kappa t}, \qquad
    \kappa t = u+\ln u .
    \label{eq:u-def-monodromy}
\end{equation}
Using Eq.~\eqref{eq:u-def-monodromy}, one obtains
\begin{align}
    z_p(\omega)
    &=
    \lim_{\epsilon\to0^+}
    \frac{C_p p}{\sqrt{2\pi}}
    \int_0^\infty
    du\,u^{p+i \omega/\kappa -1}e^{(i \omega/\kappa -\epsilon)u}
    \nonumber\\
    &=
    \frac{C_p p}{\sqrt{2\pi}}\,
    e^{i\pi p/2-\pi  \omega/(2\kappa) }
    x^{-p-i \omega/\kappa }
    \Gamma(p+i \omega/\kappa ),
    \label{eq:zp-fourier-general}
\end{align}
where the regulator makes the integral absolutely convergent
for \(\epsilon>0\), after which the limit \(\epsilon\to0^+\)
is taken. Hence
\begin{equation}
    |z_p(\omega)|^2
    =
    \frac{C_p^2p^2}{2\pi}\,
    e^{-\pi  \omega/\kappa } ( \omega/\kappa )^{-2p}
    |\Gamma(p+i \omega/\kappa )|^2 .
    \label{eq:zp-mod-general}
\end{equation}

For the present trajectory, \(p=1/2\). Using
\begin{equation}
    \left|\Gamma\left(\frac12+i \omega/\kappa \right)\right|^2
    =
    \frac{\pi}{\cosh \pi  \omega/\kappa },
\end{equation}
Eq.~\eqref{eq:zp-mod-general} gives
\begin{equation}
    |z_{1/2}(\omega)|^2
    =
    \frac{C_{1/2}^{\,2}}{4x}
    \frac{1}{e^{2\pi x}+1}.
    \label{eq:fd-monodromy-result}
\end{equation}
Substitution into Eq.~\eqref{E_spectrum_non-rel_leading}
therefore reproduces Eq.~\eqref{fd_spectrum_1}. We emphasize that the plus sign
is not a statement about fermionic field statistics; it comes from
the half-integer Lambert-\(W\) source profile together with the
Abel-defined Fourier--Mellin transform above.

For comparison, the integer case \(p=1\) gives
\begin{equation}
    |\Gamma(1+i \omega/\kappa )|^2
    =
    \frac{\pi x}{\sinh \pi  \omega/\kappa },
\end{equation}
and therefore
\begin{equation}
    |z_1(\omega)|^2
    =
    \frac{C_1^{\,2}}{ \omega/\kappa }
    \frac{1}{e^{2\pi  \omega/\kappa }-1}.
    \label{eq:be-monodromy-result}
\end{equation}
Thus, within this Lambert-\(W\) Fourier--Mellin family, integer
monodromy yields a Bose--Einstein-type denominator, whereas the half-integer monodromy of the present trajectory yields a Fermi--Dirac-type denominator.

To end this subsection, we note the following curiosity.
For real \(t\), one has \(u\in(0,\infty)\). In the Fourier
transform, the `clock' phase therefore becomes
\begin{equation}
    e^{i\omega t(u)}
    =
    e^{ix(u+\ln u)}
    =
    e^{ixu}u^{ix},
    \qquad x\equiv \frac{\omega}{\kappa}.
    \label{eq:clock-factor}
\end{equation}
The logarithm in Eq.~\eqref{eq:u-def-monodromy} carries an
imaginary-time monodromy: analytic continuation once around
\(u=0\) gives
\begin{equation}
    \ln u \longrightarrow \ln u+2\pi i,
    \qquad
    t\longrightarrow t+i\beta,
    \qquad
    \beta=\frac{2\pi}{\kappa}.
    \label{eq:imaginary-period}
\end{equation}
This is precisely the inverse-temperature scale appearing in
\(e^{\beta\omega}=e^{2\pi\omega/\kappa}\).

Under one circuit around \(u=0\),
\begin{equation}
    \left[W(e^{\kappa t})\right]^p
    =
    u^p
    \longrightarrow
    e^{2\pi i p}u^p .
    \label{eq:p-monodromy}
\end{equation}
Thus integer \(p\) has periodic monodromy, while half-integer
\(p\) has anti-periodic monodromy.
Although this (anti-)periodicity in imaginary time alone does not lead to a thermal-looking spectral distribution, it is nevertheless interesting that this property appears in classical physics.

\subsection{Energy \& Displacement in Time}
\label{sec:timepower}
To analyze the (real) temporal structure of the radiation, we use the
same variable \(u\) introduced in Eq.~(\ref{eq:u-def-monodromy}),
\be
u \equiv W(e^{\kappa t}),
\qquad\text{so that}\qquad
\kappa t = u + \ln u .
\ee
Expressing the power \(P(t)\) in terms of \(u\), and
differentiating with respect to \(u\), we find
\be
\frac{dP}{du} \propto - \frac{(3u^2 - 8u + 1)(15u^3 - 71u^2 + 33u - 1)}{(1+u)^{11}}.
\ee
Accordingly, the power's extremum is determined by the roots of the two factors in the numerator.
The cubic term determines the local maxima of three radiation bursts:
\be
15u^3 - 71u^2 + 33u - 1 = 0,
\qquad u \equiv W(e^{\kappa t_{\text{max}}}),
\ee
\begin{align}
\kappa t_{\text{max},1} &= \ln(u_1) + u_1 \approx +5.6538,\label{maxima3} \\ 
\kappa t_{\text{max},2} &= \ln(u_2) + u_2 \approx -0.2368,  \label{maxima2}\\ 
\kappa t_{\text{max},3} &= \ln(u_3) + u_3 \approx -3.3918. \label{maxima1}
\end{align}
The two local minima follow from the quadratic factor
\begin{equation}
3u^2-8u+1=0,
\qquad u\equiv W(e^{\kappa t_{\min}}),
\end{equation}
\begin{equation}
\kappa t_{\min,\pm}
=
\ln\!\left(\frac{4\pm\sqrt{13}}{3}\right)
+
\frac{4\pm\sqrt{13}}{3}.
\end{equation}
Numerically, this is:
\begin{align}
\kappa t_{\min,+} &\approx +3.4655, \\
\kappa t_{\min,-} &\approx -1.8974 .
\end{align}
Thus, although the spectrum takes a thermal Fermi-Dirac form, the power is markedly nonstationary, appearing in three distinct bursts, underscoring the kinematic, out-of-equilibrium nature of the radiation.
\begin{figure}[h] 
    \centering     \includegraphics[width=0.45\textwidth]{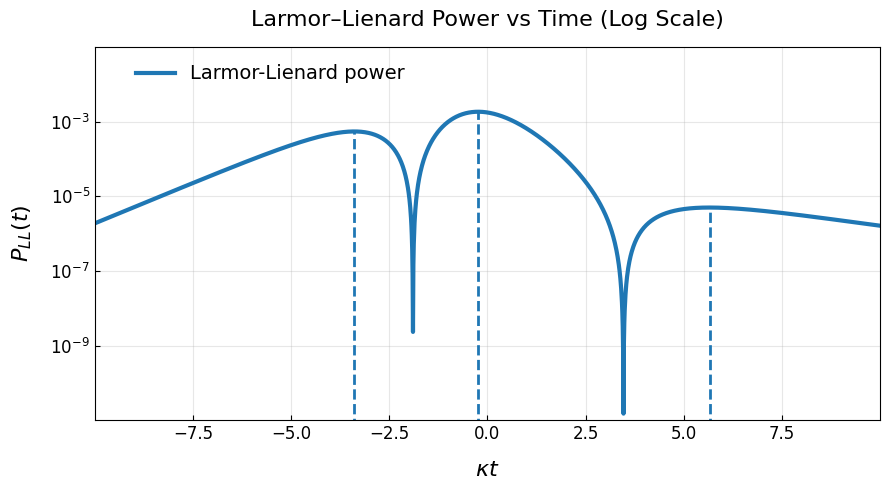}
   \caption{Instantaneous Larmor--Li\`enard radiated power $P_{\text{LL}}(t)$, Eq.~(\ref{powerLL}), plotted on a logarithmic vertical scale. The logarithmic scale makes visible the third local maximum, Eq.~(\ref{maxima3}), which is obscured on a linear vertical scale, and confirms the asymmetric triple-peak structure of the radiated power. The dashed vertical lines mark the three local maxima given in Eqs.~(\ref{maxima1}), (\ref{maxima2}), and (\ref{maxima3}).}\label{fig:power:log}
\end{figure}

\section{Spectral Information Content}
\label{sec:info}
\subsection{Single-Particle Shannon Entropy and Spectral Indistinguishability}

To further characterize the thermal-like nature of the radiated spectrum, it is instructive to evaluate its information-theoretic content in a similar spirit to, e.g., \cite{Good:2025qta, akal2021entanglement, Cong:2018vqx, Chen:2017lum, Holzhey:1994we} via an entropy measure. The Shannon differential entropy \cite{Shannon:1948continuous,CoverThomas:2006}, provides an independent measure of spectral uncertainty, decoupled from explicit energy or particle-number considerations.

Given the spectral photon number density $N(\omega)$ from Eq. (\ref{eq: number density spectrum}) and the total integrated photon number $N$ from Eq. (\ref{eq: total number of particles}), the corresponding normalized probability density $p(\omega)$ is defined as:

\begin{equation}
p(\omega) = \frac{N(\omega)}{N} = \frac{16\pi^3}{3\kappa\,\zeta(3)} \frac{(\omega/\kappa)^2}{e^{2\pi\omega/\kappa}+1}. \label{eq: probability density}
\end{equation}
The Shannon differential entropy $H$ associated with this probability distribution is governed by the standard relation\footnote{Note that this is a relation for a multiparticle distribution, not the formula for the ideal Fermi-Dirac gas.}:
$$H = -\int_0^\infty p(\omega) \ln p(\omega) \, d\omega.$$
Because $p(\omega)  d\omega$ depends solely on the dimensionless frequency ratio $x = \omega/\kappa$ and satisfies the normalization condition, the resulting entropy is inherently invariant to both the acceleration scale $\kappa$ and the global energy parameter $\hat{E}_{0}$. Evaluating this integral by executing a Fermi-Dirac series expansion combined with the Dirichlet eta function yields a scale-invariant, universal value of $H_{FD} \approx 0.0223 \text{ nats}$, which corresponds to $\approx 0.0322 \text{ bits}$ after standard base conversion. 
Note that, when written in terms of the dimensionless frequency ratio \(x=\omega/\kappa\), this Shannon entropy is independent of the acceleration scale \(\kappa\) and, consequently, of the effective temperature \(T=\kappa/(2\pi)\), as well as of the global energy parameter \(\hat{E}_0\).

The exact numerical match of the distribution Eq.~\eqref{eq: probability density} with the formal maximum-entropy Fermi--Dirac probability profile holds strictly at the level of the normalized frequency distribution, and does not imply that the emitted photons form a statistical thermal ensemble or that the physical radiation field has equilibrated.

An immediate physical consequence emerges from this correspondence: an observer restricted to the $N$-normalized one-photon-at-a-time frequency spectrum, and without access to mode-resolved occupations or multiphoton correlations, cannot distinguish the present kinematic radiation from the corresponding equilibrium Fermi--Dirac spectral
profile.
The information content encoded in this spectral context, which is inherently determined by the specific probability distribution in Eq.~(\ref{eq: probability density}), is therefore thermodynamically indistinguishable from that of equilibrium radiation. This equivalence underscores the central result of this work: thermal-like signatures and statistical-equilibrium characteristics can arise intrinsically from classical acceleration kinematics, independent of the underlying quantum field statistics or horizon physics.

\subsection{The Thermodynamic Illusion: Wien's Displacement vs. Stefan-Boltzmann Scaling}

A definitive, independent signature of the kinematic origin of this spectrum emerges when extending the thermodynamic analogy from the frequency domain to the spatial extent of the radiating source (note that this spatial characterization refers to the source trajectory itself, not to the emitted field; the electromagnetic radiation propagates freely to infinity). A $V_{\text{eff}}$ measures the effective region in which radiation was actively produced, characterizing the power-weighted variance of the charge's position along its worldline. As demonstrated in Section \ref{sec:timepower}, the spectral maximum satisfies a thermal-like Wien's displacement law, $\omega_{\textrm{max}}\propto\kappa\propto T$, where the peak frequency scales linearly with the apparent kinematic temperature $T\equiv\kappa/(2\pi)$.

However, a structural anomaly arises when analyzing the coupling between the total integrated energy and the spatial volume of the radiation zone. Integrating the spectral distribution Eq.~(\ref{eq: total emitted energy}) over all frequencies yields the total radiated energy:  
\begin{equation}
E = \hat{E}_0 \frac{7\pi \alpha}{720} T, \label{eq: integrated energy}
\end{equation}
establishing that the total integrated energy scales strictly linearly with temperature, which establishes that the total integrated energy scales strictly linearly with temperature, $E(T)\propto T$. In standard blackbody thermodynamics, a radiation field obeying either Fermi-Dirac or Bose-Einstein statistics satisfies the Stefan-Boltzmann law, wherein the spatial energy density scales as $\rho_{\textrm{thermal}}\propto T^{4}$ inside a static, externally fixed volume $V$, dictating that the total energy must scale extensively as $E\propto T^{4}$. The linear scaling $E\propto T$ in Eq.~(\ref{eq: integrated energy}) thus marks a significant departure from conventional global thermodynamic behavior, presenting an apparent problem.

This discrepancy is resolved in Appendix C, where we show that the emission zone contracts as $V_{\text{eff}}(T) \propto T^{-3}$, yielding $\rho_{\text{eff}}(T) \propto T^4$.

Consequently, the effective spatial energy density $\rho_{\textrm{eff}}\equiv E/V_{\textrm{eff}}$ within the radiation zone reduces to:
\begin{equation}
\rho_{\text{eff}}(T) = \frac{7\pi\alpha}{720\,\Lambda\sqrt{\hat{E}_0}}\, T^4.
\end{equation}

The recovery of this $T^{4}$ scaling confirms that the apparent macroscopic departure from the Stefan-Boltzmann law is purely a geometric artifact of the dynamically contracting emission zone, rather than a breakdown of analog local thermodynamic relations.

The completeness of this thermodynamic illusion becomes manifest when evaluating the remaining macroscopic parameters of the radiation zone. These derived properties are summarized in the table below:
\begin{table}[h]
\centering
\small
\begin{tabular}{|Sl|Sc|}
\hline
\textbf{Thermodynamic Quantities} & \textbf{Expressions \& Scaling Laws} \\ \hline\hline
Effective Energy Density: & $\rho_{\text{eff}}(T) = \frac{7\pi\alpha}{720\,\Lambda\sqrt{\hat{E}_0}}\, T^4 \propto T^4$ \\ \hline
Effective Heat Capacity: & $C_{\text{eff}} \equiv \frac{\partial \rho_{\text{eff}}}{\partial T} = \frac{7\pi\alpha}{180\,\Lambda\sqrt{\hat{E}_0}}\, T^3 \propto T^3$ \\ \hline
Radiation Pressure: & $P_{\text{eff}} = \frac{1}{3}\rho_{\text{eff}} = \frac{7\pi\alpha}{2160\,\Lambda\sqrt{\hat{E}_0}}\, T^4 \propto T^4$ \\ \hline
Equation of State: & $P_{\text{eff}} = \frac{1}{3}\rho_{\text{eff}} \quad \text{(Conformal State)}$ \\ \hline
Wien's Displacement Law: & $\frac{\omega_{\textrm{max}}}{T} = 3+W(3e^{-3}) = \text{const}$ \\ \hline
\end{tabular}
\end{table}

As shown, the Lambert-W kinematic trajectory not only recovers the Stefan-Boltzmann temperature scaling but mimics the entire thermodynamic landscape of standard blackbody radiation, including Debye's cubic heat capacity law and the ultra-relativistic, conformal equation of state.

\section{Conclusion}
\label{sec:conc}
We have demonstrated that the electromagnetic radiation emitted by a point charge moving along a straight line can exhibit a Fermi–Dirac spectrum, even though the underlying field is bosonic. The effect arises from a specific class of acceleration kinematics that generates out-of-equilibrium particles, yielding a closed-form spectrum with a finite total energy.

Working entirely within classical electrodynamics, we have identified the physical origin of the apparent fermionic structure as a consequence of the charge’s specific coordinate-time acceleration kinematics and its coupling to classical radiation (semi-classical particle count), rather than any appeal to quantum field theory statistics, thermal equilibrium, or horizon physics. The plus sign in the Planck denominator can be traced to the anti-periodic imaginary-time monodromy of the half-integer source profile \(W(e^{\kappa t})^{1/2}\), while an
integer monodromy instead generates the corresponding
Bose-Einstein sign. In short:
\begin{itemize}
    \item Energy spectra may look statistically thermal, but can be purely spectral-kinematic. 
    \item Seemingly equilibrium occupation formulas can emerge without equilibrium. Fermi–Dirac particle spectra need not correspond to Fermi–Dirac statistics.
    \item Temperature appearing in accelerated source radiation is thermokinematic rather than thermodynamic.
\end{itemize}
Our results demonstrate that an FD-shaped radiation spectrum, by itself, does not diagnose fermionic spin-statistics or a maximum-entropy thermal
ensemble. Rather than arising from the statistical occupation of states with equal a priori probabilities, the would-be thermality observed in the present case is of kinematic origin. Bosonic radiation may exhibit fermionic spectral structure purely as a consequence of time-dependent motion.

Eq.~\eqref{fd_spectrum_1} and Eq.~\eqref{eq: probability density}
are not equilibrium Fermi--Dirac occupation laws,
which are constrained by Pauli exclusion. Instead, because multiple
photons may occupy the same mode, the quantity
\(p(\omega)\,\diff\omega\) represents the probability that a photon
selected from the emitted radiation has frequency in the interval
\([\omega,\omega+\diff\omega]\). The remarkable feature is therefore
that bosonic radiation can carry a spectral shape, typically attributed to fermions,
without carrying their statistics.

\section{Acknowledgements} 
M.G. was supported in part by the FY2024-SGP-1-STMM Faculty Development Competitive Research Grant (FDCRGP) no.201223FD8824 and SSH20224004 at Nazarbayev University in Qazaqstan.  M.G. gives appreciation to the ROC (Taiwan) Ministry of Science and Technology (MOST), Grant no.112-2112-M-002-013, National Center for Theoretical Sciences (NCTS), and Leung Center for Cosmology and Particle Astrophysics (LeCosPA) of National Taiwan University.
The work of E.I. is supported in part by U.S. Department of Energy Grant No. de-sc0011842.

\appendix

\section{Mathematical Detail}

\subsection{Fourier integrals and related identities }
\label{sec:fourier}

Our conventions for the Fourier transform and its inverse are as follows:
\begin{equation}
    \mathcal{F}_\omega [z(t)] \equiv z(\omega)
    = \frac{1}{\sqrt{2\pi}}\int_{-\infty}^{\infty} dt\; z(t)\,e^{+i\omega t},
\label{fourier_def_1}
\end{equation}
\begin{equation}
    \mathcal{F}_t^{-1} [z(\omega)] \equiv z(t)
    = \frac{1}{\sqrt{2\pi}}\int_{-\infty}^{\infty} d\omega\; z(\omega)\,e^{-i\omega t}.
\label{fourier_def_2}
\end{equation}
Note the derivative property:
\begin{equation}
    \mathcal{F}_\omega [\dot{z}(t)] = - i \omega \mathcal{F}_\omega [z(t)].
\label{fourier_derivative_property}
\end{equation}

\subsection{Estimating the first-order correction}

Let us estimate the first-order correction to the non-relativistic approximation Eq.~\eqref{E_spectrum_non-rel_leading}.
To this end, we need to expand the exponential in Eq.~\eqref{angular_spectrum_basic_formula} while keeping a few correction terms.
As we will see in a moment, we have to keep two terms in the Taylor expansion.
We have:
\begin{equation}
\begin{aligned}
    &\int\displaylimits_{-\infty}^{\infty} \diff t\, \dot{z}(t) e^{i\omega t} e^{- i \omega z(t)\cos\theta} \\
    &\approx \int\displaylimits_{-\infty}^{\infty} \diff t\, \dot{z}(t) e^{i\omega t} \left[1 - i \omega z(t)\cos\theta + \frac{ (- i \omega z(t)\cos\theta)^2 }{2}  \right].
\end{aligned}
\end{equation}
The first term in the parentheses corresponds to the leading order (LO) approximation.
The second term (NLO) can be rewritten as, cf. \eqref{fourier_derivative_property},
\begin{equation}
    -\frac{\omega^2}{2} \cos\theta \sqrt{2\pi} \mathcal{F}_\omega [ z(t)^2 ]. 
\end{equation}
The third term (NNLO) is, analogously,
\begin{equation} 
\frac{i \omega^3 }{6} \cos^2\theta \sqrt{2\pi} \mathcal{F}_\omega [ z(t)^3 ] .
\end{equation}

Plugging this into the spectral distribution, Eq.~\eqref{angular_spectrum_basic_formula}, we arrive at:
\begin{equation}
\begin{aligned}
    &\frac{1}{\alpha} \frac{\diff I(\omega)}{\diff \Omega} \Bigg|_\text{non-rel} \approx \frac{\omega^4 \sin^2\theta }{2\pi} \left| z(\omega) \right|^2 \\
    &- \frac{\omega^5 \sin^2\theta \cos\theta }{2\pi} \mathrm{Im}\{\mathcal{F}_\omega [ z(t) ] \, \overline{\mathcal{F}_\omega [ z(t)^2 ]} \} \\
    &+ \frac{\omega^6 \sin^2\theta \cos^2\theta }{2\pi} \left\{ \frac{1}{4} \abs{\mathcal{F}_\omega [ z(t)^2 ]}^2 - \frac{1}{3} \mathrm{Re} [ \mathcal{F}_\omega [ z(t) ] \, \overline{\mathcal{F}_\omega [ z(t)^3 ] } ] \right\}.
\end{aligned}
\end{equation}
Integrating over the solid angle, we obtain the leading order energy spectrum and its corrections:
\begin{equation}
\begin{aligned}
    &E(\omega) \Bigg|_\text{non-rel} 
    \approx \frac{4 \alpha \omega^4  }{3} \left| z(\omega) \right|^2 + \\
    &+ \frac{ 4 \alpha \omega^6 }{15} \left\{ \frac{1}{4} \abs{\mathcal{F}_\omega [ z(t)^2 ]}^2 - \frac{1}{3} \mathrm{Re} [ \mathcal{F}_\omega [ z(t) ] \, \overline{\mathcal{F}_\omega [ z(t)^3 ] } ] \right\} \,.
\end{aligned}
\label{first_order_correction_formula}
\end{equation}
Note that the NLO correction integrates to zero (vanishes after angular integration), so that the first non-trivial correction comes from the NNLO term.
In a numerical treatment, for better convergence, one may use the property \eqref{fourier_derivative_property} and move some of the $\omega$'s in \eqref{first_order_correction_formula} under the Fourier transform.

Now, we estimate the correction term to obtain an upper bound. Specifically, let us take the trajectory Eq.~\eqref{z_fd_explicit}
\begin{equation}
    z(t) = \frac{ \sqrt{\hat{E}_0} }{ \kappa } \frac{\sqrt{W(e^{\kappa t})}}{ 1 + W(e^{\kappa t}) }.
\end{equation}
Here, recall that $\hat{E}_0$ parametrizes the maximum speed, see Eq.~\eqref{v_max}.
Plugging this trajectory into \eqref{first_order_correction_formula}, we obtain:
\begin{equation}
E(\omega)
\approx
\hat E_0 \frac{4\alpha}{3}
\frac{x^3}{e^{2\pi x}+1}
+
\alpha \hat E_0^2 \delta\hat E^{(1)}(x),
\qquad
x \equiv \frac{\omega}{\kappa} \,.
\label{first_order_correction_deltaEhat}
\end{equation}
The first term here, of course, is the desired spectrum from Eq.~\eqref{fd_spectrum_1}.
The correction term is parametrized by a dimensionless function $\delta \hat{E}^{(1)}$ that is independent of $\hat{E}_0$ and $\alpha$.
With this normalization, the relative correction is controlled by $\hat E_0$,
or equivalently by $v_{\max}^2$, cf. Eq.~\eqref{v_max}.
Numerically, we find that
\begin{equation}
\left|\delta\hat E^{(1)}(x)\right|
\lesssim 2.3\times 10^{-4},
\qquad x>0 .\label{correction_estimate_1}
\end{equation}
for all values of its argument.

The condition that the correction term in
Eq.~\eqref{first_order_correction_deltaEhat} be small relative
to the leading term is
\begin{equation}
\hat E_0\,
\frac{\left|\delta\hat E^{(1)}(x)\right|}
{\frac{4}{3}x^3/(e^{2\pi x}+1)}
\ll 1 ,
\end{equation}
in the frequency range where the leading spectrum is appreciable.
This condition should be imposed together with the basic
non-relativistic requirement \(v_{\max}\ll1\), namely
\begin{equation}
\hat E_0 \ll
\frac{432^2}{2(587+143\sqrt{13})}
\simeq 84.6 .
\end{equation}
Thus the first relativistic correction remains parametrically
small throughout the non-relativistic regime \(v_{\max}\ll1\).

\section{Reconstructing a generic trajectory}
\label{sec:reconstruction_recipe}

In principle, by using the same strategy, for any given reasonable spectrum $E(\omega)$, one can reconstruct a point charge trajectory $z(\omega)$ (or even a family of trajectories) that would produce electromagnetic waves with the spectrum  $E(\omega)$.
The recipe is as follows:
\begin{enumerate}
\item 
Take the desired energy spectrum $E(\omega)$.

\item 
Compute the absolute value of the Fourier-transformed trajectory as
\begin{equation}
    \abs{z(\omega)} = \sqrt{\frac{3 }{4 \alpha \omega^4} E(\omega)} \,,
\label{recipe_traj_omega}
\end{equation}
or the acceleration
\begin{equation}
    \abs{ a(\omega)} = \sqrt{\frac{3 }{4 \alpha} E(\omega)} \,.
\label{recipe_accel_omega}
\end{equation}
The latter is less singular in the limit $\omega \to 0$, making it more convenient for numerical treatment.

\item 
Solve for the trajectory by performing an inverse Fourier transform
\begin{equation}
    z(t) = \frac{1}{\sqrt{2\pi}}\int_{-\infty}^{\infty} d\omega\; |z(\omega)| e^{i \phi(\omega)} \,e^{-i\omega t} \,,
\label{recipe_traj}
\end{equation}
or find the acceleration first
\begin{equation}
    a(t) = \frac{1}{\sqrt{2\pi}}\int_{-\infty}^{\infty} d\omega\; |a(\omega)| e^{i \phi(\omega)-i\pi} \,e^{-i\omega t}
\label{recipe_accel}
\end{equation}
and then integrate it over time.
Here, $\phi(\omega)$ is the phase, which is in principle arbitrary.
The phase is restricted only by these two physically motivated conditions:
\begin{enumerate}
\item Real and smooth: $z(t)\in\mathbb{R}$ for all real $t$.
This implies Hermiticity in frequency space, $\phi(-\omega)= - \phi(\omega)$. 
\item Asymptotic inertiality: $\ddot{z}(t)\to 0$ as $t\to\pm\infty$.
This ensures finite radiated energy, $E_{\text{tot}}$, provided $\ddot{z}(t)$ is square-integrable.
\end{enumerate}

\end{enumerate}
Of course, the final step is the most likely to be intractable for a given energy spectrum $E(\omega)$.

Also, as is readily seen, the reconstruction is not unique. The trajectory is determined only up to a constant spatial shift and a constant velocity offset. The first simply reflects translational invariance, while the second affects only the \(\omega=0\) component of the non-relativistic spectral distribution, see Eq.~\eqref{specdistr_I_non-rel}. Since we are concerned only with radiation, i.e.,\ \(\omega>0\), this ambiguity is physically irrelevant.

\section{Derivation of the Effective Thermodynamic Volume}
\label{app: volume_derivation}

In this Appendix, we motivate the effective spatial volume $V_{\text{eff}}$ associated with the Lambert-W trajectory kinematics, Eq.~\eqref{eq: effective volume}, and explain how the $T^{-3}$ scaling restores Stefan-Boltzmann-like $T^4$ energy density scaling.

\subsection{Definition of $V_\text{eff}$}

We define the effective volume through the power-weighted standard deviation of the charge position:

\begin{equation}
    V_\text{eff} \sim (\Delta z)^3, \qquad \Delta z \equiv \sqrt{\langle (z - \langle z \rangle_P)^2 \rangle_P},
\end{equation}
where the power-weighted expectation value is
\begin{equation}
    \langle f \rangle_P \equiv \frac{1}{E} \int_{-\infty}^{+ \infty} f(t)\, P(t)\, dt, \qquad E = \int_{- \infty}^{+ \infty} P(t)\, dt.
\end{equation}

\subsection{Scaling from $z_{\text{max}}$}

The trajectory Eq. \eqref{z_fd_explicit} is an asymptotically resting round-trip with maximum displacement
\begin{equation}
    z_{\text{max}} = \frac{\sqrt{\hat{E}_0}}{2 \kappa} = \frac{\sqrt{\hat{E}_0}}{4 \pi T}
\end{equation}
Since the entire trajectory is confined within $|z| \leq z_{\max}$, the power-weighted spread $\Delta z$ is bounded by $z_{\max}$. Moreover, since the charge spends most of its time near $z \sim z_{\max}$ where the power is largest, we have $\Delta z \sim z_{\max}$  as an order-of-magnitude estimate:\begin{equation}
    \Delta z \;\sim\; z_{\max} \;\sim\; \frac{\sqrt{\hat{E}_0}}{\kappa} \;\propto\; \sqrt{\hat{E}_0}\, T^{-1}
\end{equation}
Cubing this characteristic scale gives
\begin{equation}
    V_{\text{eff}}(T) = \Lambda\, \hat{E}_0^{3/2}\, T^{-3}
\label{eq: effective volume}
\end{equation}
where $\Lambda$ is a dimensionless number.
The $T^{-3}$ scaling is an exact consequence of $z_{\max} \propto \kappa^{-1}$, independent of any approximation.

\subsection{Stefan-Boltzmann recovery}

The total radiated energy scales as Eq. \eqref{eq: total emitted energy}. Dividing by $V_{\text{eff}}$:
\begin{equation}
    \rho_{\text{eff}}(T) \equiv \frac{E}{V_{\text{eff}}} \propto \frac{T}{T^{-3}} = T^4.
\end{equation}
The apparent departure $E \propto T$ from the blackbody result $E \propto T^4$ is therefore a geometric artifact of the dynamically contracting emission zone: as $\kappa$ increases, $V_{\text{eff}}$ shrinks as $T^{-3}$, so the local energy density still scales as $T^4$, in agreement with the Stefan-Boltzmann law.

\newpage

\bibliography{main} 

\end{document}